Jianu, Ionuț

Conference Paper — Published Version

# The impact of government health and education expenditure on income inequality in EU

# The impact of government health and education expenditure on income inequality in European Union

**Ionuț JIANU**
Bucharest University of Economic Studies, Romania
ionutjianu91@yahoo.com

**Abstract.** *This research aims to provide an overview of the existing inequalities and their drivers in the member states of the European Union as well as their developements in the 2002-2008 and 2009-2015 sub-periods. It also analyses the impact of health and education government spending on income inequality in the European Union over the 2002-2015 period. In this context, I applied the Estimated Generalized Least Squares method using panel data for the 28-member states of the European Union.*

# 1. Introduction

One objective of the European Union agenda consists in the economic, social and territorial cohesion and recalls the higher social purpose of the European model of integration than that of the American or the Asian model. However, national social realities vary widely between the EU member states in terms of education, health, income, and employment.

The social situation does not respresents only a source for expansion of the mandate of politicians and received a special attention from the European citizens in the recent years. According to the European Commission (Reflection paper on the social dimension of Europe), 8 of 10 European citizens see unemployment, social inequalities and migration as the main challenges of the European Union. More than 50% of Europeans also believe that the next generation will be exposed to more difficult situations. In this context, the situation of social imbalances (income inequalities) across the European Union are one of the most widely debated concepts in recent years. Income inequality is natural given that people are naturaly different, have different capacities, visions, concerns, and different behaviors. For example, people's adaptability to the labour market favors high wage earnings, while rigidities in the workforce can lead to wage cuts. Rigid labour is limited in terms of the manifestation of wage discontent and, in the case of a wage cut, the employee will respect the management decision, while flexible workforce can present other employment opportunities to the management in the wage negotiation process, which can lead even to an increase of the wage of the employee by the management.

The motivation for choosing this theme consists mainly in the intensification of the economic debates on this topic, especially those related to pros and cons of income inequalities. Some researchers consider income inequality to be beneficial (given the wage disparities resulting from the different performances of the population on the labour market), while other economists support their reduction. Also, social concern regarding income gap and the methods used in order to reduce it represents an additional motivation for choosing this theme.

The objective of the paper is the examination of the situation of income inequalities in the European Union and the assesment of the impact of government health and education expenditures on income inequalities, the robustness of the assessment being strengthened by the integration of other control variables into the analysis.

# 2. Literature review

Even if many studies were made in this area, the results have sometimes been contradictory. Some economists see income inequality as an undemanding preoccupation of economists, while others consider the social sphere to be an illustration of the economic policies implemented.

According to the European Commission (2010), the socio-economic inequalities recorded in the post-2000 years were higher than those recorded in 1980, despite the economic growth achieved by the European Union. Social and human capital development policies did not have the desired impact on income inequalities due to the high labour polarization, this being a result of economic modernisation and labour market deregulation.

Eurofund (2017) analysed the evolution of pre-crisis and post-crisis incomes inequalities and found that the European Union has made significant progress in terms of convergence by 2008, but in the post-crisis period some efforts have been canceled by the impact of the economic and financial crisis that has spread stronger economic shocks in the peripheral states of the European Union than in developed member states using effective adjustment mechanisms. The impact of the crisis on the convergence process has also led to a significant decline in incomes across the European Union. The Eurofund (2017) also found that unemployment is the main channel by which the economic downturn has increased income inequality in the European Union, affecting different categories of population.

On the other hand, a series of studies (Benabou, 2000, 2002; Bleaney, Gemmell and Kneller, 2001) highlighted the pro-growth and pro-income inequality character of some public expenditure categories, such as: government expenditure on health and education, and government expenditure on infrastructure. However, other categories of expenditure may offer inadequate incentives, which implies assuming some compromises in the budget execution.

Dabla-Norris et al. (2015) identified the improvement of educational qualifications, removing financial barriers to third-country education and providing support for apprenticeship programs as factors that improve the quality of the education system and have a significant impact on income inequality. Also, in the OECD (2012) vision, educational policies that increase graduation rates in upper secondary education and in tertiary education play a fundamental role in reducing income inequality. The Organization believes that structural reforms in the labour market and those that enhance the quality of the education system are key factors in moderating income inequalities.

According to O'Donnell et al. (2013), health can influence the distribution of income through several channels, including through the one related to labour market. In this context, the labour productivity of people suffering from certain diseases is lower, which leads, generally, to lower wages. Discrimination is also another factor that can deepen income inequality between people with disabilities and healthy people. On the other hand, some researchers have also shown that the increase in income inequality is associated with high mortality (Wagstaff and Doorslaer, 2000), homicide and violence (Lynch et al., 2001). In this context, this reverse causality relationship brings significant challenges to a country in the event of an increase in income inequality, which can only be addressed by implementing effective structural reforms on the labour market (including education structural reforms).

Ward et al. (2009) demonstrated the existence of a positive relationship between the unemployment rate and the level of income inequality in the European Union. At the same time, Boltanski and Chiapiello (2005) found that labour mobility, and its adaptability, are included into the category of labour market specificities that influence income inequality. In their view, rigidity of the workforce has unfavorable consequences on the income distribution, since workers who are not willing to change their residence in order to find a better job are susceptible to commply with their modest activity and to accept lower wages.

As regards the measurement of income inequality through the Gini coefficient, Solt (2016) found that most of the statistical data sources providing this indicator is dealing with data comparability and coverage issues. In many cases, the national Gini coefficients provided by international databases are computed through different methodologies or the number of observations is low, numerous data missing from the samples. The author mentioned the only source that publishes the Gini coefficient (The Luxembourg Income Study), computed on the basis of a uniform set of definitions and assumptions, respectively on the basis of harmonised microdata that ensures maximization of their comparability. In this context, the Standardized World Income Inequality Database uses the Luxembourg Income Study standardized data series. On the other hand, the missing data was computed through the 5-year weighted moving average algorithm, the uncertainty of the results being reduced by the Monte Carlo simulation and application of the algorithm for each simulation.

## 3. Methodology

This section presents the research methodology for quantifying the impact of government health and education spending on income inequalities in the European Union. For this purpose, I used yearly data for the 2002-2015 period for each member states of the European Union (392 initial observations) as follows:

**Table 1.** *Statistical data series used*

| Indicators | Sursa datelor |
|---|---|
| Gini coefficient | The Standardized World Inequality Database and OECD for Italy (2015 year) and Luxembourg (2015 year) |
| Government expenditures on health (% of GDP) | Eurostat |
| Government expenditures on education (% of GDP) | Eurostat |
| Wages (% of GDP) | Eurostat |
| Unemployment rate (%) | Eurostat |

**Source:** Own processings using Microsoft Office Word 2016

Statistical data on income inequality, defined in this paper through Gini coefficient, are limited and the integration of this variable into an econometric model may be difficult. Eurostat does not cover the entire analysis period for this coefficient (2002-2015) and it has been necessary to consult another credible statistical platform for this indicator. I chose to use the data provided by the Standardized World Inequality Database given that the authors of the study created a new harmonised database for the Gini coefficient from several statistical sources (international studies and scientific research). Thereby, I obtained a complete set of the indicator, excepting the data for 2015 (Italy and Luxembourg - for which I used OECD data). Also, the method used maximizes the comparability of data between countries. The analysis period was limited to 2015 due to missing data for general government expenditure by function for 2016 at Eurostat level.

In the first phase I analysed the evolution of the five indicators (Table 1) in the European Union on two sub-periods: 2002-2008 and 2009-2015. The second sub-period aims to surprise the evolution of indicators during the economic crisis and the first sub-period captures the evolution of the indicators in the pre-crisis period.

In the step of identifying the influence factors and equation estimation method, I tested the stationarity of the variables through the "Summary" technique, which provides an overview of the main stationary tests: (i) Common root - Levin, Lin & Chu, (ii) Individual root - Im, Pesaran and Shin, (iii) Common root Breitung, (iv) Individual root - ADF-Fisher and (v) Individual PP-Fisher root. The variables used proved to be stationary at level and at first difference, which argued the use of the autoregressive term in the equation. Afterwards, I processed the data in Eviews 9.0 software to estimate the impact of the education and health expenditures of government on the Gini coefficient. For this purpose, I used EGLS - Estimated Generalized Least Squares method in panel window. In order to increase the feasibility of the method used, I applied the Period SUR option (for the ex-ante correction of heteroscedasticity and of the general correlations between the cross sections) on the following estimated equation:

$$\text{Gini} = \alpha_0 + \beta_0 \text{Gini}_{t-1} + \beta_1 \text{health}_{t-1} + \beta_2 \text{education}_{t-1} + \beta_3 \text{wages}_t + \beta_4 \text{un}_t + \varepsilon_t \qquad (1)$$

where: *Gini* capture income inequality, $\text{Gini}_{t-1}$ represents the autoregressive term, $\text{health}_{t-1}$ and $\text{education}_{t-1}$ surprise government health and education speding lagged by one year (expressed as a share of GDP), wages and un represents the contribution of wages to GDP, respectively the unemployment rate and $\varepsilon_t$ is the error term. Also, $\alpha_0$ represents the coefficient of the constant and $\beta_{1-4}$ captures the impact coefficients of the exogenous variables on Gini. For independent variables, I selected the lags according to the specificities of the economic theory.

Following the estimation of the equation, it was necessary to identify the method of estimating the effects. The use of autoregressive term rejects the random effects method. To confirm this hypothesis, I used Hausman test which indicated the rejection of this method. To identify the extent to which the Fixed Effect Method is appropriate for this model, I used the Redundant fixed effects test. However, dealing with issues related to multicollinearity or autocorrelation of residuals or heteroskedasticity, depending on the structure of the equation, the number of cross sections and the number of observations, may suggest using a standard panel model without fixed or random effects. Although the Fixed Effect method proved to be appropriate (Redundant fixed effects test), I rejected this technique because Period SUR option (used to correct heteroscedasticity and general cross-sectional correlations) could not be applied. I have also not been able to apply an alternative option (Cross-section SUR) given that the number of observations per cross-section (14) is less than the number of cross-sections (28) - Eviews software does not allow such an estimation.

A number of 364 observations resulted from the adjustments made for the application of this method. Further, in order to test the main assumptions for the validation of this model, I performed the following tests:
(i)   F test (verifying the statistical validity of the model);
(ii)  Jarque-Berra (examination of the normal distribution of the residuals);
(iii) Breusch-Godfrey (testing the autocorrelation of residuals);
(iv)  Breusch-Pagan and Pesaran CD (testing cross-section dependence);
(v)   Breusch-Pagan-Godfrey (testing of heteroskedasticity);
(vi)  Klein's criterion (testing for multicollinearity).

Given that the panel window do not support heteroskedasticity and autocorrelation tests, I used the theoretical framework to obtain the results of these tests. Thus, for testing the autocorrelation of residuals (iii), I estimated the following equation:

$$res1 = \gamma_0 + \delta_0 Gini_{t-1} + \delta_1 health_{t-1} + \delta_2 education_{t-1} + \delta_3 wages_t + \delta_4 un_t + res1(-1) + res1(-2) + \varepsilon_t \quad (2)$$

where: res1 represents the residuals from the initial estimated model, and res1(-1) and res1(-2) represent the series of residuals lagged by one and two years.

In order to evaluate the probability of the autocorrelation test, I used the Microsoft Office Excel 2016 function - CHISQ.DIST.RT, a function that took into account: a) the product of the R-squared corresponding to the equation (2) and the number of observations, respectively b) the degrees of freedom (number of lags for res1).

As regards, the heteroskedasticity test (v), I estimated the following equation:

$$res1\^2 = \lambda_0 + \mu_0 Gini_{t-1} + \mu_1 health_{t-1} + \mu_2 education_{t-1} + \mu_3 wages_t + \mu_4 un_t + \varepsilon_t \quad (3)$$

where res1 ^ 2 represents the square of the residuals resulting from the initial estimated model.

Finally, I used the CHISQ.DIST.RT function to evaluate the probability of the heteroskedasticity test. The function involved the following factors: a) the product of the R-squared corresponding to equation (3) and the number of observations, respectively b) the degrees of freedom (the number of independent variables, excluding the constant).

### 4. Results and interpretations

In this section I analysed the dynamics of the Gini coefficient in the European Union during 2002-2015, as well as its explanatory factors. Finally, I estimated the impact of government education and health spending and other control variables (the influence of the autoregressive term, the unemployment rate and the contribution of wages to GDP formation) on income inequality.

**Figure 1.** *Evolution of income inequality in EU-28*

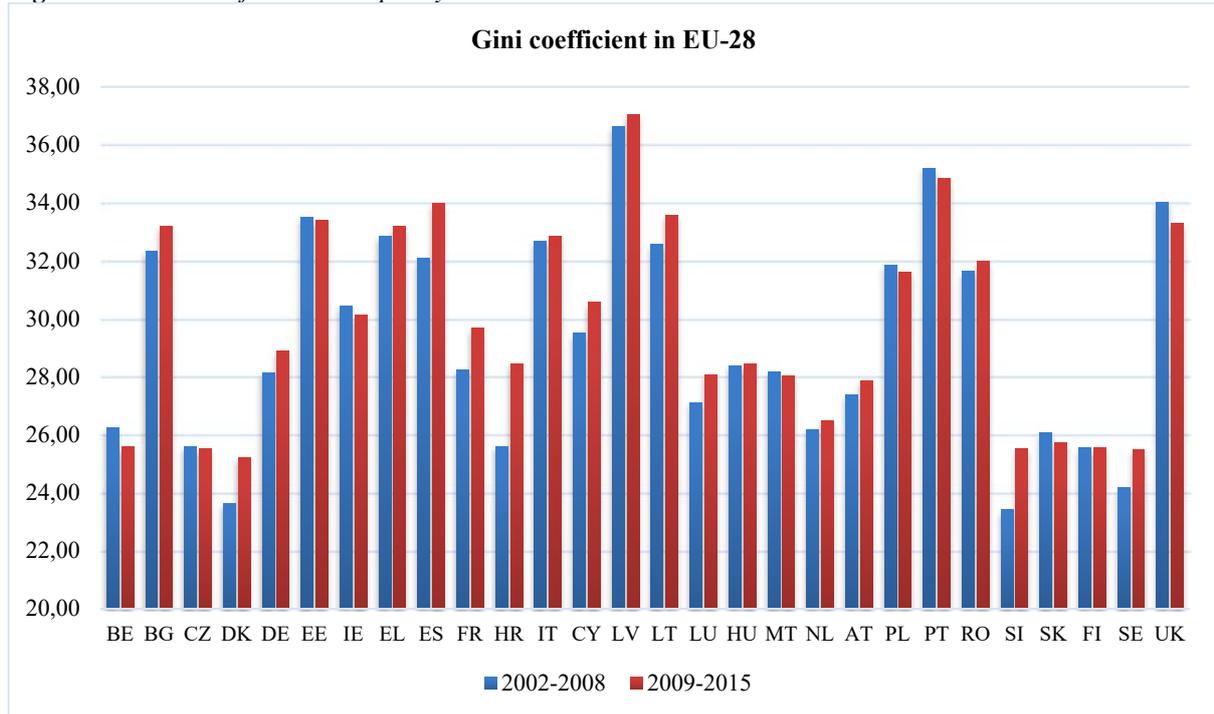

**Source:** Own calculations using Standardized World Inequality Database

Figure 1 shows that there are large differences between member states of the European Union in terms of income inequality. According to the Standardized World Inequality Database, in the post-crisis period, the European Union experienced a trend of increasing income inequalities (at the end of 2015, the European Union average of the Gini coefficient was 30.0). Also, I would like to mention that for all indicators analysed, I have processed their final value at European Union level, taking into account the year of accession of each member state to European Union. In this context, the average Gini coefficient for the 2009-2015 sub-period increased by 0.67 deviation points compared to the one recorded in the previous sub-period (from 29.16 to 29.83). In 2015, the highest Gini coefficient was recorded in Latvia (36.8), Portugal (34.8) and Spain (34.3), while the lowest levels of the indicator have been observed in Denmark (25.4), Czech Republic (25.5), Slovakia (25.5) and Finland (25.5). In the 2009-2015 sub-period, only 9 states recorded decreases in the indicator compared to the 2002-2008 sub-period, while 19 experienced increases. The most significant increases in income inequality were recorded in Croatia (2.86 deviation points), Slovenia (2.11 deviation points) ) and Spain (1.90 deviation points). On the other hand, all the 9 countries that have experienced cuts in the indicator in the second sub-period recorded insignificant decreases, below 1 deviation points. As can be seen, in most of the EU member states, the crisis had a negative impact on the income gap.

**Table 2.** *The commitment to reducing inequality rank (152 countries)*

| Country | Commitment to reducing inequality | Health, education and social protection expenditure | Progressive structure and incidence of tax | Labour market policies to address inequality |
|---|---|---|---|---|
| Sweden | 1 | 9 | 8 | 8 |
| Belgium | 2 | 4 | 3 | 24 |
| Denmark | 3 | 8 | 9 | 12 |
| Germany | 5 | 2 | 17 | 6 |
| Finland | 6 | 3 | 23 | 10 |
| Austria | 7 | 6 | 40 | 1 |
| France | 8 | 5 | 19 | 21 |
| The Netherlands | 9 | 19 | 13 | 9 |
| Luxembourg | 10 | 12 | 21 | 11 |
| Ireland | 13 | 1 | 53 | 19 |
| Italy | 16 | 17 | 14 | 29 |

| Country | Commitment to reducing inequality | Health, education and social protection expenditure | Progressive structure and incidence of tax | Labour market policies to address inequality |
|---|---|---|---|---|
| United Kingdom | 17 | 28 | 31 | 5 |
| Portugal | 19 | 18 | 29 | 30 |
| Slovenia | 20 | 13 | 56 | 22 |
| Malta | 22 | 37 | 2 | 26 |
| Czech Republic | 24 | 10 | 104 | 14 |
| Greece | 25 | 11 | 60 | 46 |
| Spain | 27 | 16 | 48 | 55 |
| Hungary | 28 | 21 | 85 | 32 |
| Cypru | 31 | 42 | 38 | 27 |
| Slovakia | 32 | 23 | 128 | 20 |
| Croatia | 33 | 44 | 32 | 39 |
| Poland | 35 | 22 | 121 | 38 |
| Estonia | 38 | 26 | 127 | 43 |
| Latvia | 46 | 31 | 145 | 28 |
| Romania | 50 | 57 | 132 | 31 |
| Bulgaria | 79 | 52 | 144 | 44 |
| Lithuania | 83 | 49 | 141 | 49 |

**Source**: Own processings using data from Development Finance International and Oxfam „The Commitment to Reducing Inequality Index", July 2017

However, the inclusive feature of the European integration model is also reflected in the positions occupied by the member states of the European Union in a ranking made by the Development Finance International in partnership with the international confederation of charities - Oxfam (the inequality reduction commitment index). As can be seen, the ranking (Table 2) is in line with the situation of income inequalities (Gini) in the member states of the European Union, the countries recording high levels of Gini index being also among the countries that make insufficient efforts to reduce income inequalities. The first countries in the world that make full efforts to reduce income inequality through government spending on health, education and social protection, progressive taxation and tax incidence, respectively labour market policies are Sweden, Belgium and Finland. On the other hand, Romania, Bulgaria and Lithuania occupy the last three places in the European Union from the point of view of the comittment to reduce income inequality.

The economic crisis has also played a decisive role in the dynamics of income inequality. According to Eurostat, in 2009, the EU-28 GDP fell by 4.3%, evolution accompanied by a large budget deficit of 6.6% of GDP, which required the adoption of fiscal consolidation policies by the member states. Although member states adopted austerity policies, many of these has not reduced the share of government spending on education and health in GDP. Annex 1 highlights the developments at European level in the 2002-2008 and 2009-2015 sub-periods regarding government expenditure on education and health, unemployment rate and wages. As regards government education expenditures in European Union, their share in GDP of 4.84% in 2015 was the lowest value in the analysed period. However, the average of the indicator for the 2009-2015 sub-period was higher than the one recorded in the previous sub-period by 0,08% of GDP. In 2015, Romania has recorded the lowest government spending on education (% of GDP), followed by Ireland, Bulgaria and Italy. On the other hand, the member states of the European Union spending the most on education are: Denmark, Sweden, Belgium and Finland. In the 2009-2015 sub-period, only 8 countries reduced the share of government education spending in GDP compared to the values recorded in the 2002-2008 sub-period, the most significant being: Hungary (-0.71% of GDP), Romania (-0.54% of GDP) and Poland (-0.53% of GDP). The most significant increases in education spending were found in Denmark (+0.63% of GDP), Luxembourg (+0.54% of GDP) and Belgium (+0.53% of GDP).

The dynamic of government health expenditure highlighted the same preference of member states for rising expenditure in the second sub-period, with only 5 countries of European

Union opting for a decrease in this budgetary function: Hungary (-0.37% of GDP), Portugal (-0.27% of GDP), Greece (-0.26% of GDP), Bulgaria and Malta (both -0.13% of GDP). This preference of member states for increasing government health expenditure also facilitated an increase of the aggregate indicator at European Union level by 0.7% of GDP in the 2009-2015 sub-period, compared to the value recorded in the previous sub-period. The most significant increases in the second sub-period analysed were recorded in the Netherlands (+2.04% of GDP), Finland (+1.31% of GDP) and the UK (+1.31 of GDP). In 2015, Denmark, France and the Netherlands had the highest government health spending, while Cyprus, Latvia and Romania made the lowest government health spending from the European Union.

Some countries have laid the foundations of fiscal consolidation on reducing government social expenditures and thus, the income gap has further increased. However, unemployment had a strong influence on income discrepancies. Even if the unemployment rate at EU level in the analysed period has reached a maximum level in 2013 (10.93%), its level is still high (9.42% in 2015) and poses an important challenge for the European Union as a whole. In the post-crisis period (2009-2015), the EU average unemployment rate was higher than in the previous period by 1.82 percentage points (from 8.08% to 9.90%).

According to Eurostat, in 2015 the highest unemployment in the European Union was recorded in Greece (24.9%), Spain (22.1%) and Croatia (16.1%), while the lowest level of this indicator was found in Germany (4.6%), Czech Republic (5.1%) and the United Kingdom (5.3%). Only in seven EU member states unemployment rate declined in the 2009-2015 sub-period compared to the previous one. The largest decline was found in Poland (-6.14 percentage points), followed by Germany (-3.61 percentage points) and Slovakia (-1.77 percentage points). On the other hand, the highest increases in unemployment were recorded in Spain (+12.21 percentage points), Greece (+11.11 percentage points) and Ireland (+8.23 percentage points). At first glance, it may be said that the rise in unemployment has led to the jobs loss of individuals with higher incomes too, and income inequality should not have changed. However, people who lost their jobs and previously earned better than other individuals, have been able to use their savings to generate a substitute income for their wage achieved in the past.

Unemployment also affected vulnerable groups, especially the people who have attained the International Standard Classification of Education taxonomic classes - ISCED 0-2 (less than primary, primary and lower secondary education) and ISCED 3-4 (upper secondary and post-secondary non-tertiary education). In this context, the EU-28 unemployment rate of the 15-74 age group who attained ISCED 0-2 increased in 2009 from the level recorded in 2008 by 3.2 percentage points, reaching 14.4%. On the other hand, the unemployment rate of the 15-74 age group with ISCED level 3-4 in the EU-28, increased by 1.8 percentage points, from 6.5% in 2008 to 8.3 % in 2009. Regarding the population with tertiary education (ISCED 5-8), the economic recession led to an increase in the unemployment rate by 1.1 percentage points in 2009 (4.9%), compared to the rate recorded in the previous year, which also argues the higher resilience to economic shocks of these categories of people.

As regards the share of wages in GDP, the indicator was constant during the analysed period and fluctuated around 37% of GDP. However, there were large discrepancies between the indicator recorded at the level of each member state. As a result, in 2015, the highest share of wages in GDP was recorded in Denmark (47.9% of GDP), Slovenia (41.7% of GDP) and Luxembourg (41.5% of GDP) while Ireland (24.8% of GDP), Greece (25.0% of GDP) and Romania (27.3% of GDP) had the weakest position of the indicator. In the second sub-period analysed, the largest increases of the contribution of wages to GDP formation compared to the previous sub-period were found in Bulgaria (+5.11% of GDP), Cyprus (+2.74% of GDP) and Finland (+2.56% of GDP). Contrariwise, Romania (-3.37% of GDP), Ireland (-2.43% of GDP) and Portugal (-1.87% of GDP), recorded the largest decline in the average of the indicator processed for the sub-period 2009-2015, compared to the previous sub-period.

As the increasing evolution of inequalities poses new challenges to the social dimension of the European Union, in the second phase of the research, I assessed the impact of the drivers of income inequalities on the Gini coefficient. Initially, I checked the stationarity of the variables included in the model (using the Summary method mentioned in the methodology - the lag being automatically chosen by Eviews software using the Schwarz information criterion), these being stationary at level and first difference, which required the inclusion of the Gini coefficient lagged by 1 year in the regression. Therefore, following the examination of the stationarity tests, I obtained the following results:

- Gini coefficient - stationarity identified at I(1);
- government expenditure on health as a share of GDP - stationarity identified at I(1);
- government expenditure on education as a share of GDP - stationarity identified at I(0);
- unemployment rate - the use of the Summary method did not provide a clear picture of the results (the number of tests that identified stationarity at I(0) was equal to the number of tests that identified stationarity at I(1)); Consequently, I applied the Hadri test for both level and first difference of unemployment rate. I rejected the stationarity assumption for I(0) as the probability of 0% was below the 5% threshold. The probability of 56% associated with Hadri Z-stat for I(1) argued the acceptance of the null hypothesis of stationarity. Finally, I have identified stationarity at I(0);
- contribution of wages to GDP formation - stationarity identified at I(1).

Next, I estimated the model, starting from the structure previously presented and I analysed its results. According to Figure 2, all variables used are statistically significant, their probability being less than the significance threshold of 5%. However, the risk of the estimator of the constant to be null is greater than 5% and less than 10% (8.21%), but this does not raise any questions with respect to the appropriate representation of the model given that the coefficient of determination (R-squared) is high, which demonstrates that the dynamic of the selected independent variables explains 99.72% of the dependent variable variance. Also, the low standard errors have increased the confidence in estimators. In order to check the validity of the model, I used the F-test and its associated probability created the premises for confiming the statistical validity of the model.

The analysis of the impact coefficients of the variables included in the model is based on the "caeteris-paribus" hypothesis. According to the results of the model attached in Figure 3, the increase of the Gini coefficient recorded in the previous year by 1 deviation point leads to an increase of the actual Gini coefficient by 0.989 deviation points, this being also caused by the higher yield of higher incomes (people who earn higher income than other categories of people may use these additional resources to generate other types of income, a situation that lead to an increase in income gap).

Returning to the main focus of the analysis, I identified a negative impact of the increase in government education and health expenditure on the Gini coefficient in the European Union. Thus, I found that an increase in government health expenditure in the previous year by 1% of GDP leads to a decrease by 0.019 deviation points in the Gini coefficient, a lower impact than the one manifested through government education expenditure channel (an increase of 1% of GDP in government education expenditure lagged by one year led to a decline of 0.024 deviation points in Gini coefficient). Education offers to population the opportunity to alling with people who earn high income, within certain limits, through knowledge. Supporting education through higher funding, (one condition of it being the effectiveness of the structural reforms) can stimulate young people's insertion on labour market or their involvement in sustainable projects. However, the effects of the investments in education are not observed on the short term. This budget function generates new sources of income for the population by offering opportunities on the medium or long term, reason for which I included the 1-year lagged series on the list of the independent variables. This is also the case of government health expenditure that support education over time. In general, there is a significant link between education and health, since an educated person tends to take care of his health condition and a healthy person has the ability to continue his studies, not being constrained by factors related to weak health condition. Also, a healthy person does not face health barriers in the accession on the labour market and its performances are not conditioned or limited by the current condition of health. However, this is not a rule, the career success also depending on the severity of the health problem and other specific psychological factors.

According to Figure 2, an increase of the contribution of wages to the GDP formation by 1% of GDP leads to an increase of the Gini coefficient by 0.007 deviation points. This effect can be caused by the fact that in many countries from the European Union wage hikes occurred unequally, favouring the population earning high wages.

The estimate shows a positive impact of 0.011 deviation points on the Gini coefficient at a 1 percentage point increase in the unemployment rate, based on the higher impact of unemployment on vulnerable groups.

The coefficient of the constant term shows that, when all components of the equation remain constant, the Gini coefficient increases by 0.268 deviation points. However, given that the level of R-squared is high and the risk of the estimator to be null is greater than 5%, I ignored its coefficient as it can be inaccurate.

**Figure 2.** *Results of the model*

```
Dependent Variable: GINI
Method: Panel EGLS (Period SUR)
Sample (adjusted): 2003 2015
Periods included: 13
Cross-sections included: 28
Total panel (balanced) observations: 364
Linear estimation after one-step weighting matrix
```

| Variable | Coefficient | Std. Error | t-Statistic | Prob. |
|---|---|---|---|---|
| GINI(-1) | 0.988523 | 0.003412 | 289.6816 | 0.0000 |
| HEALTH(-1) | -0.019482 | 0.007470 | -2.607897 | 0.0095 |
| EDUCATION(-1) | -0.023601 | 0.010690 | -2.207825 | 0.0279 |
| WAGES | 0.007654 | 0.002276 | 3.362277 | 0.0009 |
| UN | 0.011262 | 0.002475 | 4.549447 | 0.0000 |
| C | 0.268036 | 0.153712 | 1.743752 | 0.0821 |

Weighted Statistics

| | | | |
|---|---|---|---|
| R-squared | 0.997218 | Mean dependent var | 126.3769 |
| Adjusted R-squared | 0.997179 | S.D. dependent var | 94.80012 |
| S.E. of regression | 0.996300 | Sum squared resid | 355.3557 |
| F-statistic | 25661.19 | Durbin-Watson stat | 1.996223 |
| Prob(F-statistic) | 0.000000 | | |

Unweighted Statistics

| | | | |
|---|---|---|---|
| R-squared | 0.995730 | Mean dependent var | 29.55577 |
| Sum squared resid | 19.62285 | Durbin-Watson stat | 0.898298 |

**Source:** Own calculations using Eviews 9.0

As can be seen in Figure 3, the residuals are normally distributed, given that the probability of Jarque-Bera test is higher than 5% (52.46%).

In order to test the autocorrelation of the residuals, I did not use the Durbin-Watson test as it became invalid when I introduced the autoregressive term, as an exogenous variable, in the panel model. In this context, I performed the Breusch-Godfrey test (Annex 2), starting from equation (2) which includes 2 degrees of freedom (the number of lags for residuals) and a number of 308 observations (following the adjustments performed) and I obtained a R-squared value of 0.012849 and a n*R-square value of 3.957586. Using the CHISQ.DIST.RT function and previously processed data, I computed the probability of the Breusch-Godfrey autocorrelation test (13.82%), which argued the acceptance of the hypothesis according to which there is not autocorrelation between the residuals.

For enhancing the examination of the model's accuracy, I performed the cross-section dependence tests: Breusch-Pagan and Pesaran CD. Both the Breusch-Pagan (100.00%) and the Pesaran CD test (63.76%) are superior to the significance threshold of 5%, which confirms the absence of cross-section dependence.

The verification of heteroskedasticity (Annex 3) involved the estimation of the equation (3) and the computation of the probability of Breusch-Pagan-Godfrey test based on the n*R-squared value (364 * 0.012786 = 4.6541049) and the degrees of freedom taken into consideration (5 - number of exogenous variables). Therefore, I accepted the hypothesis of homoskedasticity since the Breusch-Pagan-Godfrey probability of 45.95% is higher than 5%.

**Figure 3.** *Distribution of the residuals*

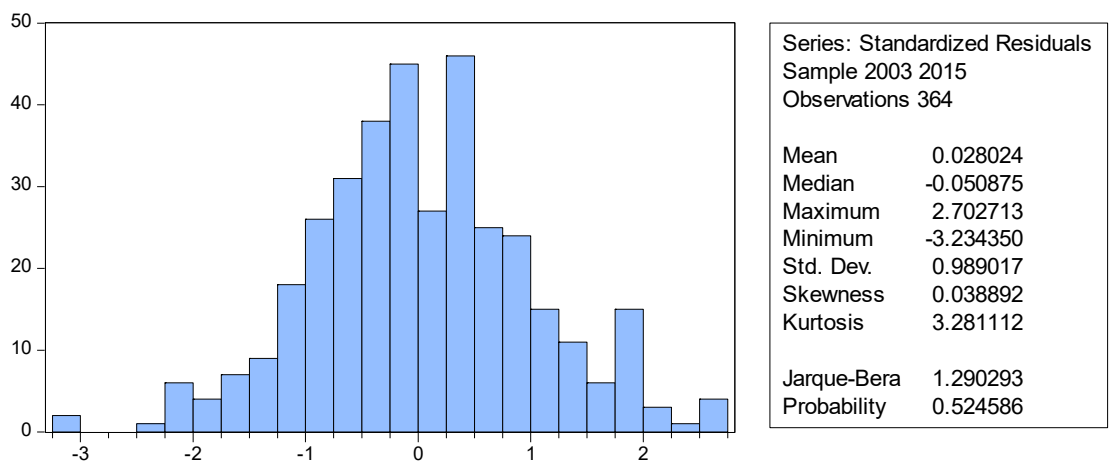

**Source:** Own calculations using Eviews 9.0

Concerning multicoliniarity, I accepted the hypothesis related to its absence from the model, since the Pearson statistical correlations between the exogenous variables are lower than the coefficient of determination of the equation (1) - the Klein criterion.

Finally, I validated the model and its coefficients, given that there is no reason to have doubts on the maximum verisimilitude of the estimators.

## 5. Conclusions

This paper targeted the estimation of the impact of government health and education expenditure on income inequality. According to the results, a 1 percentage point increase in government health expenditure (express as a share of GDP) leads to a reduction in the Gini coefficient by 0.019 deviation points in the next year, while the same dynamic of government education expenditure causes a decrease of Gini coefficient by 0.024 deviation points in the next year. Therefore, this analysis confirms the inverse relationship between these two functions of budget expenditures and income inequality. The analysis of the EU member states' commitments to reduce inequalities has also confirmed this hypothesis. The model has proved to be statistically valid, all the tests performed in order to confirm the maximum verisimilitude of the estimators providing results that were in the normal parameters.

Supporting social cohesion through government spending on education and health is essential, but it is necessary to make it more effective by implementing structural reforms that bring both social and economic benefits outweighing the budgetary costs resulting from the implementation of these measures. Spending public money inefficiently can have negative consequences on the living standards of future generations, given that at some point the population will have to comply with their obligations resulted from high public debt, as a consequence of large budget deficits.

At both European Union and Romanian level, it is necessary to identify an optimal threshold of income inequality, a lower level than it - representing the natural income inequality and a higher level than the threshold - being the inequality induced by the national institutions and governments. In this context, for European Union member states it would be beneficial to assess the impact of budgetary measures on income distribution in the context of the annual budget proposals and to prevent the accumulation of disparities between incomes (for instance, Finland have such an approach, even if this activity is not mandated by law). Otherwise, there is a risk that the actual level of income inequality will continue on a increasing trend.

**Annex 1.** *The dynamic of* government *health and education expenditure, wage share in GDP and unemployment rate in the EU-28*

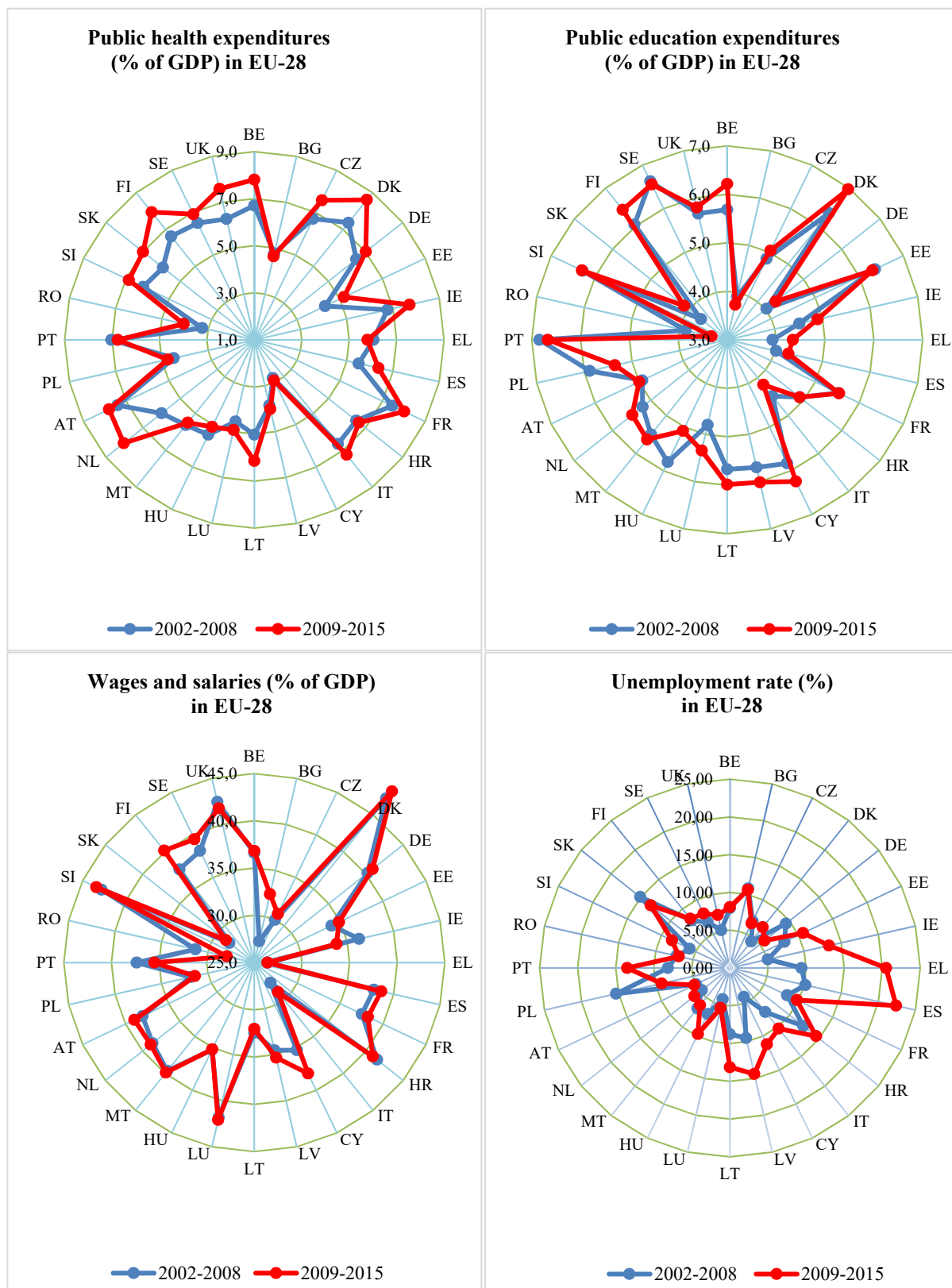

**Source:** Own calculations using Eurostat database

**Annex 2.** *Autocorrelation of the residuals test - Breusch-Godfrey*

Dependent Variable: RESID01
Method: Panel EGLS (Period SUR)
Sample (adjusted): 2005 2015
Periods included: 11
Cross-sections included: 28
Total panel (balanced) observations: 308
Linear estimation after one-step weighting matrix

| Variable | Coefficient | Std. Error | t-Statistic | Prob. |
|---|---|---|---|---|
| GINI(-1) | 0.002436 | 0.020108 | 0.121149 | 0.9037 |
| HEALTH(-1) | 0.013366 | 0.045458 | 0.294031 | 0.7689 |
| EDUCATION(-1) | 0.011475 | 0.069190 | 0.165854 | 0.8684 |
| WAGES | 0.002688 | 0.014205 | 0.189244 | 0.8500 |
| UN | -0.024565 | 0.014500 | -1.694139 | 0.0913 |
| C | -0.063632 | 0.903669 | -0.070415 | 0.9439 |
| RESID01(-1) | -0.006266 | 0.056106 | -0.111683 | 0.9111 |
| RESID01(-2) | 0.008319 | 0.056097 | 0.148303 | 0.8822 |

Weighted Statistics

| | | | |
|---|---|---|---|
| R-squared | 0.012849 | Mean dependent var | 0.016518 |
| Adjusted R-squared | -0.010184 | S.D. dependent var | 1.007006 |
| S.E. of regression | 1.011935 | Sum squared resid | 307.2035 |
| F-statistic | 0.557853 | Durbin-Watson stat | 1.998322 |
| Prob(F-statistic) | 0.789857 | | |

Unweighted Statistics

| | | | |
|---|---|---|---|
| R-squared | 0.006179 | Mean dependent var | 0.022359 |
| Sum squared resid | 297.3885 | Durbin-Watson stat | 1.995710 |

**Source:** Own calculations using Eviews 9.0

**Annex 3.** *Heteroskedasticity test - Breusch-Pagan-Godfrey*

Dependent Variable: RESID01^2
Method: Panel EGLS (Period SUR)
Sample (adjusted): 2003 2015
Periods included: 13
Cross-sections included: 28
Total panel (balanced) observations: 364
Linear estimation after one-step weighting matrix

| Variable | Coefficient | Std. Error | t-Statistic | Prob. |
|---|---|---|---|---|
| GINI(-1) | 0.014939 | 0.018358 | 0.813753 | 0.4163 |
| HEALTH(-1) | 0.003934 | 0.045262 | 0.086920 | 0.9308 |
| EDUCATION(-1) | -0.068923 | 0.065222 | -1.056748 | 0.2913 |
| WAGES | 0.026728 | 0.013392 | 1.995764 | 0.0467 |
| UN | 0.016312 | 0.015968 | 1.021556 | 0.3077 |
| C | -0.252463 | 0.860992 | -0.293224 | 0.7695 |

Weighted Statistics

| | | | |
|---|---|---|---|
| R-squared | 0.012786 | Mean dependent var | 0.813186 |
| Adjusted R-squared | -0.001002 | S.D. dependent var | 1.089608 |
| S.E. of regression | 1.004172 | Sum squared resid | 360.9932 |
| F-statistic | 0.927335 | Durbin-Watson stat | 1.988804 |
| Prob(F-statistic) | 0.463184 | | |

Unweighted Statistics

| | | | |
|---|---|---|---|
| R-squared | -0.002726 | Mean dependent var | 0.976252 |
| Sum squared resid | 794.8883 | Durbin-Watson stat | 1.761169 |

**Source:** Own calculations using Eviews 9.0